\documentclass{ws-p9-75x6-50}

\begin{document}

\title{15-60 Hz QPO and Periastron Precession around the X-ray  Neutron Star
Magnetosphere}

\author{ C.M. Zhang}

\address{IFT, UNESP, Rua Pamplona 145, 01405-900, Sao Paulo,
Brasil\\E-mail: zhangcm@ift.unesp.br}

\maketitle

\abstracts{
Based on the periastron precession  model to account for kHz 
QPO of the binary X-ray neutron star, 
proposed by Stella and Vietri, we ascribe the 
15-60 Hz Quasi Periodic Oscillation (QPO) to the periastron
precession frequency of the orbiting accreted matter 
  at the boundary of magnetosphere-disk of X-ray neutron star (NS). 
 The obtained conclusions include: all QPO frequencies increase  
with increasing the
accretion rate. The  theoretical relations between 15-60 Hz QPO (HBO) 
frequency 
and the  twin kHz QPOs are  similar to the measured  
empirical formula. Further,  the better fitted 
 NS mass by the proposed model  is about 1.9 solar masses for the
detected LMXBs. }

\def\be{\begin{equation}}
\def\ee{\end{equation}}
\def\bea{\begin{eqnarray}}
\def\eea{\end{eqnarray}}
\def\c{\cite}

\def\et{ {\it et al.}}
\def\la{ \langle}
\def\ra{ \rangle}
\def\ov{ \over}

\def\mdot{\ifmmode \dot M \else $\dot M$\fi}    
\def\mxd{\ifmmode \dot {M}_{x} \else $\dot {M}_{x}$\fi}
\def\med{\ifmmode \dot {M}_{Edd} \else $\dot {M}_{Edd}$\fi}
\def\bff{\ifmmode B_{f} \else $B_{f}$\fi}

\def\apj{\ifmmode Astrophys. J. \else Astrophys. J.\fi}    
\def\apjl{\ifmmode Astrophys. J. Lett.
 \else Astrophys. J. Lett.\fi}    %
\def\aap{\ifmmode Astron. and Astrophys.
 \else Astron. and Astrophys.\fi}    %
\def\mnras{\ifmmode Mon. Not. Royal  Astron. Society 
 \else Mon. Not. Royal  Astron. Society \fi}    %
\def\aaps{\ifmmode Astron. and Astrophys. Suppl.
 \else Astron. and Astrophys. Suppl.\fi}    %
\def\nat{\ifmmode Nature (London)\else Nature (London)\fi}
\def\prl{\ifmmode Phys. Rev. Lett. \else Phys. Rev. Lett.\fi}
\def\prd{\ifmmode Phys. Rev. D. \else Phys. Rev. D.\fi}

\def\ms{\ifmmode M_{\odot} \else $M_{\odot}$\fi}    
\def\no{\ifmmode \nu_{1} \else $\nu_{1}$\fi}    
\def\nt{\ifmmode \nu_{2} \else $\nu_{2}$\fi}    
\def\ntmax{\ifmmode \nu_{2max} \else $\nu_{2max}$\fi}    
\def\nomax{\ifmmode \nu_{1max} \else $\nu_{1max}$\fi}    
\def\nh{\ifmmode \nu_{\rm HBO} \else $\nu_{\rm HBO}$\fi}    
\def\nqpo{\ifmmode \nu_{QPO} \else $\nu_{QPO}$\fi}    
\def\nz{\ifmmode \nu_{o} \else $\nu_{o}$\fi}    
\def\nht{\ifmmode \nu_{H2} \else $\nu_{H2}$\fi}    
\def\ns{\ifmmode \nu_{s} \else $\nu_{s}$\fi}    
\def\nb{\ifmmode \nu_{burst} \else $\nu_{burst}$\fi}    
\def\nkm{\ifmmode \nu_{km} \else $\nu_{km}$\fi}    
\def\dn{\ifmmode \Delta\nu \else $\Delta\nu$\fi}    

\def\rs{\ifmmode R_{s} \else $R_{s}$\fi}    
\def\rmm{\ifmmode R_{M} \else $R_{M}$\fi}    
\def\rco{\ifmmode R_{co} \else $R_{co}$\fi}    
\def\rim{\ifmmode R_{I} \else $R_{I}$\fi}    

\section{The Model}
On the kHz QPO mechanism\c{k20}, recently, the  general
relativistic periastron precession effects are paid much attetion 
to account for kHz QPOs proposed by Stella and
Vietri\c{sv99}, which can explain the varied kHz QPOs separation
$\dn$. Moreover, HBO frequency\c{hk89} ($\nh \simeq 15 - 60 $Hz), 
 is interpreted to be the beat frequency between
the Keperian frequency of the magnetosphere-disk  and the stellar
spin frequency \c{as85}.
Later, it was considered to be  
 the nodal precession \c{vs98} of Lense-Thirring
effect in the disk. In the  model of SV99 \c{sv99}, 
the twin kHz QPOs are ascribed to the Keperian
frequency 
and the periastron 
frequency of material orbiting the neutron star at some
disk radius, i.e., 
$\nu_{2} = \nu_K = (M/4\pi^2 r^3)^{1/2}$
 and $\nu_{1} = \nu_{K}[1 -  (1 - 6M/r)^{1/2}]\;$, where r is the
Schwarzschild coordinate distance and M is the gravitational mass 
of neutron star (We set the unity of the speed of light and the
gravitational constant $c=G=1$).
Here,  we concentrate on the explanation of  HBOs
(for Atoll sources,  15 - 60 Hz QPO is supposed to be
the same mechanism as HBO of  Z sources \c{psa99} and
its relation to the twin kHz QPOs.
 We assume  that $\nh$  is a periastron  precession
frequency  of the accreted orbiting  materials  
in the magnetosphere-disk boundary, and the twin kHz QPOs are 
that proposed  in SV99\c{sv99} are produeced in the inner disk. 
There exists a scaling factor to connect two radii, which 
will be determined by the well fitted  data of kHz QPO and HBO.
 Therefore, these basic frequencies are written as follows through 
defining the suitable parameter y, which is the ratio between the 
Schwarzschild radius to the disk radius, 
\be
\no (y) = (1-\sqrt{1-3y})\times {\; }  \nt (y)\;, 
\label{no}
\ee
\be
\nt (y) = {\nz}  y^{3/2}\;,
\;\;\;\nz \equiv {11300\over m}{\;}{(Hz)}\;,
\label{nt}
\ee
\be
\nh (y) =  {\nz} (1-\sqrt{1-3\phi y})\times    (\phi y)^{3/2}{\; }
\;,\;\;\; \phi \equiv {r \over R_{M}}\;, 
\label{nh}
\ee
 where  $R_{M}$ is the radius 
 of NS  magnetosphere, which is inversely related to the accretion rate 
 and proportionally related to the magnetic field strength of the star.
 $\phi$ is a scaling parameter to connect the radii of the inner disk and 
 NS  magnetosphere, but here we suppose it to be $0.3 \sim 0.4 $
 for the reason of the best fitting.

The relations $\nh$ vs.  $\nt$  is plotted in 
Fig.1, together with the well measured five Z-source samples, 
and it is shown that the agreement  between the model and 
the observed QPO
data is quite well.

\begin{figure}[t]
\vskip -8.5cm
\epsfxsize=20pc
\epsfbox{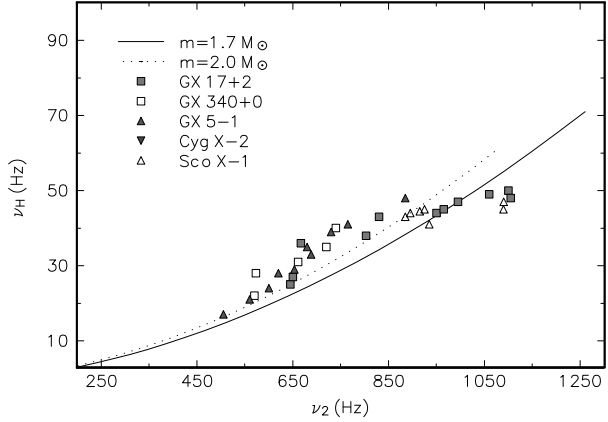}
\caption[fig1.eps]
{ HBO frequency  versus the upper kHz QPO
frequency  for five Z sources of  LMXBs \c{psa99}. 
Error bars are not plotted for the
sake of clarity.
The model presents a well fitting for  the  nearly circular
orbit of NS mass about 2.0 solar mass with the scaling parameter 
$\phi = 0.4$.}
\end{figure}

 From Eqs.(\ref{no}), (\ref{nt}) and (\ref{nh}),
we can derive the theoretical
relations between QPO frequencies   in the
following,
\be
\nh \simeq 50.6 {\;}(Hz){\;}({\no\over 500})
[ 1 - 0.15 ({m\no \over 500})^{2/5} ] \;,
\label{nhno}
\ee
\be
\nh \simeq 30.4 {\;}(Hz){\;}m^{2/3}({\nt\over 1000})^{5/3}
[1 + 0.07 ({m\nt\over 1000})^{2/3}] \;,
\label{nhnt}
\ee
\be
\no \simeq 300{\;}(Hz){\;}m^{2/3}({\nt\over 1000})^{5/3}
[1 + 0.2({m\nt\over 1000})^{2/3}] \;.
\label{nont}
\ee
These theoretical relations are consistent with the empirical 
relations in ref.[6]. 


\begin{thebibliography}{99}
\bibitem{k20}
M., van der Klis,  2000, submitted (astro-ph/0001167).
\bibitem{sv99}L., Stella,  and M., Vietri,  \prl,  {\bf 82}, 17
(1999).
\bibitem{hk89}
G., Hasinger, and M., Van der Klis, \aap, {\bf 225}, 79 (1989).
\bibitem{as85}
A. Alpar, \& J. Shaham, \nat,  {\bf 316}, 239 (1985);
F.K. Lamb, N. Shibazaki, A. Alpar, \& J. Shaham, \nat, {\bf 317}, 681
(1985).
\bibitem{vs98}M.,  Vietri,  \& L., Stella,  \apj, {\bf 503}, 350
(1998).
\bibitem{psa99}D., Psaltis,  {\it et al}, \apj,
 {\bf 520}, 262 (1999);  \apj, {\bf 501}, L95 (1998).
\end{thebibliography}
\end{document}